\documentstyle[prd,preprint,aps]{revtex}
\begin{document}
\tightenlines
\draft
\title{Detecting the MSSM Higgs Bosons at Future $e^+e^-$ Colliders}

\author{A. Guti\'errez-Rodr\'{\i}guez $^{1}$, M. A. Hern\'andez-Ru\'{\i}z $^{2}$ and O. A. Sampayo $^{3}$}

\address{(1) Escuela de F\'{\i}sica, Universidad Aut\'onoma de Zacatecas\\
          Apartado Postal C-580, 98060 Zacatecas, Zacatecas M\'exico.}

\address{(2) Facultad de Matem\'aticas, Universidad Aut\'onoma de Zacatecas\\
          Apartado Postal C-612, 98060 Zacatecas, Zacatecas M\'exico.}

\address{(3) Departamento de F\'{\i}sica, Universidad Nacional del Mar del Plata\\
          Funes 3350, (7600) Mar del Plata,  Argentina.}

\date{\today}
\maketitle
\begin{abstract}
We investigate the possibility of detecting the Higgs bosons predicted in
the Minimal Supersymmetric extension of the Standard Model $(h^0, H^0, A^0, H^\pm)$,
with the reactions $e^{+}e^{-}\rightarrow b\bar b h^0 (H^0, A^0)$, and
$e^+e^-\to \tau^-\bar \nu_\tau H^+, \tau^+\nu_\tau H^-$, using the helicity
formalism. We analyze the region of parameter space $(m_{A^0}-\tan\beta)$
where $h^0, H^0, A^0$ and $H^\pm$ could be detected in the limit when $\tan\beta$
is large. The numerical computation is done considering two stages of a possible 
Next Linear $e^{+}e^{-}$ Collider: the first with $\sqrt{s}=500$ $GeV$
and design luminosity 50 $fb^{-1}$, and the second with $\sqrt{s}=1$ $TeV$ and
luminosity 100-200 $fb^{-1}$.
\end{abstract}

\narrowtext
\section{Introduction}
Higgs bosons \cite{Higgs} play an imporant role in the Standard Model (SM) \cite{Weinberg};
they are responsible for generating the masses of all the elementary particles
(leptons, quarks, and gauge bosons). However, the Higgs-boson sector is the
least tested one in the SM. If Higgs bosons are responsible for breaking the
symmetry from $SU(2)_L\times U(1)_Y$ to $U(1)_{EM}$, it is natural to expect
that other Higgs bosons are also involved in breaking other symmetries at the
grand-unification scale. One of the more attractive extensions of the SM is
Supersymmetry (SUSY) \cite{Nilles}, mainly because of its capacity to solve
the naturalness and hierarchy problems while maintaining the Higgs bosons
elementary.

The minimal supersymmetric extension of the Standard Model (MSSM) doubles the
spectrum of particles of the SM and the new free parameters obey simple
relations. The scalar sector of the MSSM \cite{Gunion} requires two Higgs doublets,
thus the remaining scalar spectrum contains the  following physical states:
two CP-even Higgs scalar ($h^0$ and $H^0$) with $m_{h^0}\leq m_{H^0}$,
one CP-odd Higgs scalar ($A^0$) and a charged Higgs pair ($H^{\pm}$), whose
detection would be a clear signal of new physics. The Higgs sector is
specified at tree-level by fixing two parameters, which can be chosen as the
mass of the pseudoscalar $m_{A^0}$ and the ratio of vacuum expectation
values of the two doublets $\tan \beta = v_{2}/v_{1}$, then the mass
$m_{h^0}$, $m_{H^0}$ and $m_{H^{\pm}}$ and the mixing angle of the neutral
Higgs sector $\alpha$ can be fixed. However, since radiative corrections
produce substantial effects on the predictions of the model \cite{Li}, it is
necessary to specify also the squark masses, which are assumed to be
degenerated. In this paper, we focus on the phenomenology of the neutral
CP-even and CP-odd scalar ($h^0, H^0, A^0$) and charged $(H^\pm)$.  

The search for these scalars has begun at LEP, and current low energy bound
on their masses gives $m_{h^0}$, $m_{A^0}$ $>$ 90 $GeV$ and $m_{H^\pm}$
$>$ 120 $GeV$ for $\tan\beta$ $>$ 1 \cite{Review}.
At $e^{+}e^{-}$ colliders the signals for Higgs bosons are relatively clean
and the opportunities for discovery and detailed study will be excellent. The
most important processes for the production and detection of the neutral and
charged Higgs bosons $h^0$, $H^0$, $A^0$ and $H^\pm$, are: $e^{+}e^{-}\rightarrow
Z^{*}\rightarrow h^0, H^0+Z^0$, $e^{+}e^{-}\rightarrow Z^{*} \rightarrow
h^0, H^0+A^0$, $e^{+}e^{-}\rightarrow \nu \bar \nu + W^{+*}W^{-*}
\rightarrow \nu \bar \nu+h^0, H^0$ (the later is conventionally referred
to as $WW$ fusions), and $e^+e^-\to H^+H^-$ \cite{Komamiya}; precise cross-section formulas
appear in Ref. \cite{Gunion1}. The main decay modes of the neutral Higgs particles
are in general $b\bar b$ decays $(\sim 90 \%)$ and $\tau^+\tau^-$ decays
$(\sim 10 \%)$ which are easy to detect experimentally at $e^+e^-$
colliders \cite{Janot,Carena,Sopczak}. Charged Higgs particles decay predominantly into
$\tau \nu_\tau$ and $t\bar b$ pairs.

The $Z^0h^0$ production cross-section contains an overall factor $\sin^{2}(\beta-\alpha)$
which suppress it in certain parameter regions (with $m_{A^0} < 100$ $GeV$
and $\tan\beta$ large); fortunately the $A^0h^0$ production cross-section contains
the complementary factor $\cos^{2}(\beta-\alpha)$. Hence the $Z^0h^0$ and $A^0h^0$
channels together are well suited to cover all regions in the $(m_{A^0}-\tan\beta)$
plane, provided that the $c.m.$ energy is high enough for $Z^0h^0$ to be produced through
the whole $m_{h^0}$ mass range, and that an adequate event rate can be achieved. 
These conditions are already shown to be satisfied \cite{Janot,Carena,Sopczak} for $\sqrt{s}= 500$ $GeV$
with assumed luminosity $10$ $fb^{-1}$, as is expected to be the case of the Next Linear $e^{+}e^{-}$
Collider (NLC).

In previous studies, the two-body processes $e^{+}e^{-}\rightarrow h^0(H^0)+Z^0$
and $e^{+}e^{-}\rightarrow h^0(H^0)+A^0$ have been evaluated \cite{Gunion1} extensively.
However, the inclusion of three-body process $e^{+}e^{-}\rightarrow h^0(H^0)+b\bar b$
and $e^{+}e^{-}\rightarrow A^0+b\bar b$ \cite{Cotti} at future $e^+e^-$ colliders energies \cite{NLC,NLC1,JLC}
is necessary in order to know its impact on the two-body mode processes and
also to search for new relations that could have a cleaner signature of the Higgs
bosons production.

\noindent In the other hand the decay modes of the charged Higgs bosons determine
the signatures in the detector. If $m_{H^\pm}> m_t + m_b$, the dominant decays modes are $t\bar b$,
$\bar t b$ and $\tau^+\nu_\tau$, $\tau^-\bar \nu_\tau$. In some part of the
parameter space also the decay $H^+\to W^+h^0$ is allowed. If $m_{H^\pm} < m_t + m_b$,
the charged Higgs boson will decay mainly into $\tau^+\nu_\tau$, $\tau^-\bar \nu_\tau$.

\noindent For $m_{A^0}\leq m_{Z^0}$, and if 50 events criterion are adecuate, the
$H^+H^-$ pair production will be kinematically allowed and easily observable
\cite{Janot,Carena,Sopczak,Djouadi2,Brignole,Gunion4}. For $m_{A^0}>120$ $GeV$,
$e^+e^- \rightarrow H^+H^-$ must be employed for detection of the three heavy
Higgs bosons. Assuming that SUSY decays are not dominant, and using the 50 event
criterion $H^+H^-$ can be detected up to $m_{H^\pm}=230$ $GeV$
\cite{Janot,Carena,Sopczak,Djouadi2,Brignole,Gunion4}, assuming $\sqrt{s}=500$ $GeV$.

\noindent The upper limits in the $H^+H^-$ mode are almost entirely a function
of the machine energy (assuming an appropriately higher integrated luminosity
is available at a higher $\sqrt{s}$). Two recent studies \cite{Gunion5,Feng}
show that at $\sqrt{s}=1$ $TeV$, with an integrated luminosity of 200 $fb^{-1}$,
$H^+H^-$ detection would extended to $m_{A^0}\sim m_{H^\pm} \sim 450$ $GeV$
even if substantial SUSY decays of these heavier Higgs are present.

In the present paper we study the production of SUSY Higgs bosons at $e^{+}e^{-}$ colliders.
We are interested in finding regions that could allow the detection of the SUSY
Higgs bosons for the set parameter space $(m_{A^0}-\tan\beta)$. We shall discuss
the neutral and charged Higgs bosons production $b\bar b h^0 (H^0, A^0)$, and
$\tau^-\bar \nu_\tau H^+$, $\tau^+\nu_\tau H^-$ in the energy range of a future
$e^+e^-$ colliders \cite{NLC,NLC1,JLC} for large values of the parameter
$\tan\beta$, where one expects to have a high production. Since the coupling
$h^0b\bar b$ is proportional to $\sin\alpha/\cos\beta$, the cross-section will
receive a large enhancement factor when $\tan\beta$ is large. Similar situation
occurs for $H^0$, whose coupling with $b\bar b$ is proportional to $\cos\alpha/\cos\beta$.
The couplings of $A^0$ with $b\bar b$ and of $H^\pm$ with $\tau^-\bar \nu_\tau, \tau^+\nu_\tau$
are directly proportional to $\tan\beta$, thus the amplitudes will always grow with
$\tan\beta$. We consider the complete set of Feynman diagramas at tree-level
and use the helicity formalism \cite{Howard,Zhan,Werle,Pilkum,Peter,Mangano,Berends}
for the evaluation of the amplitudes. The results obtained for the three-body
processes are compared with the dominant two-body mode reactions for the plane
$(m_{A^0}-\tan\beta)$. Succintly, our aim in this work is to analyze how much
the results of the Bjorken Mechanism [Fig. 1, (1.4)] would be enhanced by the
contribution from the diagrams depicted in Figs. 1.1-1.3, 1.5 and 1.6 in which
the SUSY Higgs bosons are radiated by a $b (\bar b)$ quark. For the case of the
charged Higgs bosons the two-body mode [Figs. 3.1 and 3.4] would be enhanced by
the contribution from the diagrams depicted in Figs. 3.2, 3.3, and 3.5,
in which the charged Higgs boson is radiated by a $\tau^- \bar \nu_\tau$
$(\tau^+ \nu_\tau)$ lepton.

Recently, it has been shown that for large values of $\tan\beta$ the detection
of SUSY Higgs bosons is possible at FNAL and LHC \cite{Lorenzo}. In the papers
cited in Ref. \cite{Lorenzo} the authors calculated the corresponding three-body diagrams
for hadron collisions. They pointed out the importance of a large bottom
Yukawa coupling at hadron colliders and showed that the Tevatron collider
may be a good place for detecting SUSY Higgs bosons. In the case of the
hadron colliders the three-body diagrams come from gluon fusion and this
fact makes the contribution from these diagrams more important, due to the
gluon abundance inside the hadrons. The advantage
for the case of $e^{+} e^{-}$ colliders is that the signals of the processes
are cleaner.

This paper is organized as follows. We present in Sec. II the relevant details
of the calculations. Sec. III contains the results for the processes
$e^{+}e^{-}\rightarrow b\bar bh^0 (H^0,A^0)$ and $e^+e^-\to \tau^-\bar \nu_\tau H^+,
\tau^+ \nu_\tau H^-$ at future $e^+e^-$ colliders. Finally, Sec. IV contains
our conclusions.

\section{Helicity Amplitude for Higgs Bosons Production}

When the number of Feynman diagrams is increased, the calculation of the amplitude
is a rather unpleasant task. Some algebraic forms \cite{Hearn} can be used in it to
avoid manual calculation, but sometimes the lengthy printed output from the computer
is overwhelming, and one can hardly find the required results from it. The CALKUL 
collaboration \cite{Causmaecker} suggested the Helicity Amplitude Method (HAM) which can
simplify the calculation remarkably and hence make the manual calculation realistic.

In this section we describe in brief the evaluation of the amplitudes at tree-level, for
$e^{+}e^{-}\rightarrow b\bar b h^0 (H^0, A^0)$ and $e^+e^-\to \tau^-\bar \nu_\tau H^+,
\tau^+ \nu_\tau H^-$   using the HAM \cite{Howard,Zhan,Werle,Pilkum,Peter,Mangano,Berends}.
This method is a powerful technique for computing helicity amplitudes for multiparticle
processes involving massles spin-1/2 and spin-1 particles. Generalization of this method 
that incorporates massive spin-1/2 and spin-1 particles, is given in Ref. \cite{Berends}.
This algebra is easy to program and more efficient than computing the Dirac algebra.

A Higgs boson $h^0, H^0$, $A^0$, and $H^\pm$ can be produced in scattering $e^{+}e^{-}$ via
the following processes:

\begin{eqnarray}
e^{+}e^{-} &\rightarrow& b\bar b h^0,\\
e^{+}e^{-} &\rightarrow& b\bar b H^0,\\
e^{+}e^{-} &\rightarrow& b\bar b A^0,\\         
e^+e^-&\rightarrow& \tau^-\bar \nu_\tau H^+, \tau^+ \nu_\tau H^-.
\end{eqnarray}

The diagrams of Feynman, which contribute at tree-level to the different reaction
mechanisms, are depicted in Figs. 1-3. Using the Feynman rules given by the
Minimal Supersymmetric Standard Model (MSSM), as summarized in Ref. \cite{Gunion1},
we can write the amplitudes for these reactions \cite{Cotti}. For the evaluation
of the amplitudes we have used the spinor-helicity technique of Xu, Zhang and
Chang \cite{Zhan} (denoted henceforth by XZC), which is a modification of the
technique developed by the CALKUL collaboration \cite{Causmaecker}. Following
XZC, we introduce a very useful notation for the calculation of the processes
(1)-(3) \cite{Cotti} and (4). The complete formulas of the processes (1)-(3)
are given in Ref. \cite {Cotti}. For the case of process (4), we present the
relevant details of the calculations.

Let us consider the process

\begin{equation}
e^{-}(p_{1}) + e^{+}(p_{2}) \rightarrow \{\tau^-(k_{2}) + 
\bar \nu_{\tau}(k_{3}) + H^+(k_{1}), 
\tau^+(k_{2}) + 
 \nu_{\tau}(k_{3}) + H^-(k_{1})\},
\end{equation}

\noindent in which the helicity amplitude is denoted by 
${\cal M}[\lambda (e^{-}), \lambda (e^{+}), \lambda (\tau^{\mp}), 
\lambda (\nu_\tau)]$.

Due to charge invariance, the cross-sections for the production processes
$e^-e^+ \to \tau^-\bar \nu_{\tau}H^+$ and $e^-e^+ \to \tau^+\nu_{\tau}H^-$
are exactly the same. One should thus calculate the cross-section of one of
the two reactions and then multiply by a factor of two to take into account
the charge conjugate final state. This will enormously simplify the Feynman
diagrams as well as the amplitudes of transition. The Feynman diagrams for
this process are shown in Fig. 3. From this figure it follows that the amplitudes
that correspond to each graph are

\begin{eqnarray}
{\cal M}_1&=&-iC_{1}P_{H^-}(k_2+k_3)P_{Z}(p_1+p_2)T_1,\nonumber\\
{\cal M}_2&=&iC_{2}P_{\tau}(k_1+k_3)P_{Z}(p_1+p_2)T_2,\nonumber\\
{\cal M}_3&=&-iC_{3}P_{\nu}(k_1+k_2)P_{Z}(p_1+p_2)T_3,\\
{\cal M}_4&=&-iC_{4}P_{H^-}(k_2+k_3)P_{\gamma}(p_1+p_2)T_4,\nonumber\\
{\cal M}_5&=&iC_{5}P_{\tau}(k_1+k_3)P_{\gamma}(p_1+p_2)T_5,\nonumber
\end{eqnarray}

\noindent where

\begin{eqnarray}
C_{1}&=&-\frac{g^3}{16\sqrt2}\frac{m_\tau}{m_W}\tan\beta
\frac{\cos2\theta_W}{\cos^2\theta_W}, \nonumber\\
C_{2}&=&\frac{g^3}{32\sqrt2}\frac{m_\tau}{m_W}\tan\beta
\frac{1}{\cos^2\theta_W}, \nonumber\\
C_{3}&=&C_2, \\
C_{4}&=&\frac{g^3}{2\sqrt2}\frac{m_\tau}{m_W}\tan\beta
\sin^2\theta_W, \nonumber\\
C_{5}&=&C_4, \nonumber
\end{eqnarray}

\noindent while the propagators are

\begin{eqnarray}
P_{Z^0}(p_1+p_2)&=&\frac{(s-m_{Z^0}^2)+i m_{Z^0} \Gamma_{Z^0}}
{(s-m_{Z^0}^2)^2+(m_{Z^0} \Gamma_{Z^0})^2},\nonumber\\
P_{H^\pm}(k_2+k_3)&=&\frac{(2k_2\cdot k_3-m_{H^\pm}^2)+im_H\Gamma_{H^\pm}}
{(2k_2\cdot k_3-m_{H^\pm}^2)^2+(m_{H^\pm}\Gamma_{H^\pm})^2},\nonumber\\
P_{\tau}(k_1+k_3)&=&\frac{1}{m_{H^\pm}^2+2k_1\cdot k_3},\\
P_{\nu}(k_1+k_2)&=&\frac{1}{m_{H^\pm}^2+2k_1\cdot k_2},\nonumber\\
P_{\gamma}(p_1+p_2)&=&\frac{1}{s},\nonumber
\end{eqnarray}

\noindent where $s=(p_1+p_2)^2$ and the corresponding tensors are

\begin{eqnarray}
T^{\mu}_1&=&\bar u(k_2)(1-\gamma_5)v(k_3)\bar v(p_2)
(k\llap{/}_{1}-k\llap{/}_{2}-k\llap{/}_{3})(v^z_e-a^z_e\gamma_5)u(p_1)
,\nonumber\\
T^{\mu}_2&=&\bar u(k_2) \gamma^{\mu} (v^z_e-a^z_e\gamma_5)
(k\llap{/}_{1}+k\llap{/}_{3})(1-\gamma_5)v(k_3)
\bar v(p_2)\gamma_{\mu}(v^z_e-a^z_e\gamma_5)u(p_1)
,\nonumber\\
T^{\mu}_3&=&\bar u(k_2)(1-\gamma_5)(k\llap{/}_{1}+k\llap{/}_{2})\gamma_{\mu}
(v^z_\nu-a^z_\nu\gamma_5)v(k_3)\bar v(p_2)\gamma^{\mu}
(v^z_e-a^z_e\gamma_5)u(p_1)
,\\
T^{\mu}_4&=&\bar u(k_2)(1-\gamma_5)v(k_3)\bar v(p_2)
(k\llap{/}_{1}-k\llap{/}_{2}-k\llap{/}_{3})u(p_1)
,\nonumber\\
T^{\mu}_5&=&\bar u(k_2)\gamma_{\mu}(k\llap{/}_{1}+k\llap{/}_{3})
(1-\gamma_5)v(k_3)\bar v(p_2)\gamma^{\mu}u(p_1)
.\nonumber
\end{eqnarray}

In fact, we rearrange the tensors $T^{'}$s in such a way that they become appropriate to
a computer program. Then, following the rules from helicity calculus formalism
\cite{Howard,Zhan,Werle,Pilkum,Peter,Mangano,Berends} and using identities of the type

\begin{equation}
\{\bar u_{\lambda}(p_{1})\gamma ^{\mu}u_{\lambda}(p_{2})\}\gamma_{\mu}=2u_{\lambda}(p_{2})\bar u_{\lambda}(p_{1})+2u_{-\lambda}(p_{1})\bar u_{-\lambda}(p_{2}),
\end{equation}

\noindent which is in fact the so called Chisholm identity, and

\begin{equation}
p\llap{/}=u_{\lambda}(p)\bar u_{\lambda}(p)+u_{-\lambda}(p)\bar u_{-\lambda}(p),
\end{equation}

\noindent defined as a sum of the two projections $u_{\lambda}(p)\bar u_{\lambda}(p)$
and $u_{-\lambda}(p)\bar u_{-\lambda}(p)$.

The spinor products are given by

\begin{eqnarray}
s(p_{i}, p_{j})&\equiv&\bar u_{+}(p_{i})u_{-}(p_{j})=-s(p_{j}, p_{i}),\nonumber\\
t(p_{i}, p_{j})&\equiv&\bar u_{-}(p_{i})u_{+}(p_{j})=[s(p_{j}, p_{i})]^{*}.
\end{eqnarray}

\noindent Using Eqs. (10)-(12), which are proved in Ref. \cite{Berends}, we
can reduce many amplitudes to expressions involving only spinor products.

Evaluating the tensors of Eq. (9) for each combination of  $(\lambda, \lambda ^{'})$
with $\lambda, \lambda^{'} =\pm 1$ one obtains the following expressions:

\begin{eqnarray}
{\cal M}_{1}(+,+)&=&F_{1}f_{1}^{+,+}s(k_{2},k_{3})
[s(p_2,k_1)t(k_1,p_1)-s(p_2,k_2)t(k_2,p_1)-s(p_2,k_3)t(k_3,p_1)],\nonumber\\
{\cal M}_{1}(-,+)&=&F_{1}f_{1}^{-,+}s(k_{2},k_{3})
[t(p_2,k_1)s(k_1,p_1)-t(p_2,k_2)s(k_2,p_1)-t(p_2,k_3)s(k_3,p_1)],\\
{\cal M}_2(+,+)&=&F_2 f_2^{+,+} s(k_2,p_2)t(p_1,k_1)s(k_1,k_3),\nonumber\\
{\cal M}_2(-,+)&=&F_2 f_2^{-,+} s(k_2,p_1)t(p_2,k_1)s(k_1,k_3),\\
{\cal M}_3(+,+)&=&F_3 f_3^{+,+} s(k_2,k_1)t(k_1,p_1)s(p_2,k_3),\nonumber\\
{\cal M}_3(-,+)&=&F_3 f_3^{-,+} s(k_2,k_1)t(k_1,p_2)s(p_1,k_3),\\
{\cal M}_{4}(+,+)&=&F_{4}s(k_{2},k_{3})
[s(p_2,k_1)t(k_1,p_1)-s(p_2,k_2)t(k_2,p_1)-s(p_2,k_3)t(k_3,p_1)],\nonumber\\
{\cal M}_{4}(-,+)&=&F_{4}s(k_{2},k_{3})
[t(p_2,k_1)s(k_1,p_1)-t(p_2,k_2)s(k_2,p_1)-t(p_2,k_3)s(k_3,p_1)],\\
{\cal M}_5(+,+)&=&F_5  s(k_2,p_2)t(p_1,k_1)s(k_1,k_3),\nonumber\\
{\cal M}_5(-,+)&=&F_5  s(k_2,p_1)t(p_2,k_1)s(k_1,k_3),
\end{eqnarray}

\noindent where

\begin{eqnarray}
F_{1}&=&-2iC_{1}P_{H^\pm}(k_2+k_3)P_{Z^0}(p_1+p_2),\nonumber\\
F_{2}&=&4iC_{2}P_{\tau}(k_1+k_3)P_{Z^0}(p_1+p_2),\nonumber\\
F_{3}&=&-4iC_{3}P_{\nu}(k_1+k_2)P_{Z^0}(p_1+p_2),\\
F_{4}&=&-2iC_{4}P_{H^\pm}(k_2+k_3)P_{\gamma}(p_1+p_2),\nonumber\\
F_{5}&=&4iC_{5}P_{\tau}(k_1+k_3)P_{\gamma}(p_1+p_2),\nonumber
\end{eqnarray}

\noindent and

\begin{eqnarray}
f_{1}^{+,+}&=&(v^z_e-a^z_e),\nonumber\\
f_{1}^{-,+}&=&(v^z_e+a^z_e),\nonumber\\
f_{2}^{+,+}&=&(v^z_e-a^z_e)^2,\nonumber\\
f_{2}^{-,+}&=&((v^z_e)^2-(a^z_e)^2),\nonumber\\
f_{3}^{+,+}&=&(v^z_\nu+a^z_\nu)(v^z_e-a^z_e),\nonumber\\
f_{3}^{-,+}&=&(v^z_\nu+a^z_\nu)(v^z_e+a^z_e).\nonumber
\end{eqnarray}

\noindent Here, $v^z_{e}=-1+4\sin ^{2}\theta _{W}$, $a^z_{e}=-1$, 
$v^z_{\nu}=1$ and $a^z_{\nu}=1$, according to the experimental data \cite{Review}. 

After the evaluation of the amplitudes of the corresponding diagrams, we obtain
the cross-sections of the analyzed processes for each point of the phase space
using Eqs. (13)-(17) by a computer program, which makes use of the subroutine
RAMBO (Random Momenta Beautifully Organized). The advantages of this procedure
in comparison to the traditional ``trace technique" are discussed in Refs.
\cite{Howard,Zhan,Werle,Pilkum,Peter,Mangano,Berends}.

We use the Breit-Wigner propagators for the $Z^{0}$, $h^0$, $H^0$, $A^0$ and
$H^{\pm}$ bosons. The mass of the bottom $(m_b \approx 4.5 $GeV$)$ the mass
$(M_{Z^0} = 91.2 $GeV$)$ and width $(\Gamma_{Z^0} = 2.4974 $GeV$)$ of $Z^{0}$
have been taken as inputs; the widths of $h^0$, $H^0$, $A^0$ and $H^{\pm}$ are
calculated from the formulas given in Ref. \cite{Gunion1}. In the next sections
we present the numerical computation of the processes $e^+e^-\to b\bar bh$,
$h=h^0, H^0, A^0$ and $e^+e^-\to \tau^-\bar \nu_{\tau} H^+, \tau^+\nu_\tau H^-$.

\section{Detection of MSSM Higgs Bosons at Future Positro-Electron Colliders Energies}

In this paper, we study the detection of neutral and charged MSSM Higgs bosons at
$e^{+}e^{-}$ colliders, including three-body mode diagrams [Figs. 1.1-1.3, 
1.5, and 1.6; Figs. 2.1-2.3, 2.5 and 2.6; Figs. 3.2, 3.3, and 3.5]
besides the dominant mode diagram [Fig. 1.4; Fig. 2.4; Figs. 3.1, and 3.4]
consider two stages of a possible Next Linear $e^{+}e^{-}$ Collider: the first
with $\sqrt{s}=500$ $GeV$ and design luminosity 50 $fb^{-1}$, and the second with
$\sqrt{s}=1$ $TeV$ and luminosity 100-200 $fb^{-1}$. We consider the complete set
of Feynman diagrams (Figs. 1-3) at tree-level and utilize the helicity formalism
for the evaluation of their amplitudes. In the next subsections, we present our
results for the case of the different Higgs bosons.

\subsection{Detection of $h^0$}

In order to illustrate our results on the detection of the $h^0$ Higgs boson, 
we present graphs in the parameters space region $(m_{A^0}-\tan\beta)$, assuming 
$m_{t}= 175$ $GeV$, $M_{\stackrel{\sim}t}= 500$ $GeV$ and $\tan\beta > 1$ for
NLC. Our results are displayed in Fig. 4, for $e^{+}e^{-}\rightarrow (A^0, Z^0)+ h^0$
dominant mode and for the processes at three-body $e^{+}e^{-}\rightarrow b\bar b h^0$.

The total cross-section for each contour is $0.03$ $pb$, and $0.01$ $pb$, which
gives 1500 events, and 500 events to an integrated luminosity of ${\cal L}=50$
$fb^{-1}$. We can see from this figure, that the effect of the reaction $b\bar b h^0$
is not more important that $(A^0, Z^0) + h^0$, for most of the $(m_{A^0}-\tan\beta)$
parameter space regions. Nevertheless, there are substantial portions of parameter
space in which the discovery of the $h^0$ is not possible using either $(A^0, Z^0) + h^0$
or $b\bar b h^0$.

For the case of $\sqrt{s}=1$ $TeV$, the results of the detection of the $h^0$
are shown in Fig. 5. It is clear from this figure that the contribution of the
process $e^{+}e^{-}\rightarrow b\bar b h^0$ becomes dominant, namely
$e^{+}e^{-}\rightarrow (A^0, Z^0) + h^0$ is small in all parameter space. However,
they could provide important information on the Higgs bosons detection. For instance,
we give the contours for the total cross-section to, say 0.01 $pb$, 0.005, and
0.003 $pb$ for both processes. These cross-sections give 1000 events, 500 events,
and 300 events in total to a integrated luminosity of ${\cal L}=100$ $fb^{-1}$. While for
${\cal L}= 200$ $fb^{-1}$ the events number is 2000, 1000, and 600, respectively,
then it will be detectable the $h^0$ at future $e^+e^-$ colliders.

\subsection{Detection of $H^0$}

To illustrate our results regarding the detection of the heavy Higgs bosons $H^0$,
we give the contours for the total cross-section for both processes
$e^{+}e^{-}\rightarrow (A^0, Z^0) + H^0$, $e^{+}e^{-}\rightarrow b\bar b H^0$
in Fig. 6 for $\sqrt{s}= 500$ $GeV$ and ${\cal L} = 50$ $fb^{-1}$. The contours 
for this cross-section are 0.01 $pb$, 0.001 $pb$ and 0.0001 $pb$ for both reactions
$(A^0, Z^0) + H^0$ and $b\bar b H^0$. The number of events corresponding
to each contour are 500, 50 and 5, respectively.

Our estimate is that if more than 100 total events are obtained for a given process
$(A^0, Z^0) + H^0 \hspace*{2mm} or \hspace*{2mm}b\bar b H^0$ then the Higgs boson
$H^0$ can be detectable.

For the case of $\sqrt{s}=1$ $TeV$, the results on the detection of the $H^0$ are
show in Fig. 7. The events number for each contour is 1000, 100, and 10 for
${\cal L}=100$ $fb^{-1}$ and 2000, 200, 20 for ${\cal L}=200$ $fb^{-1}$.

The effect of incorporate $b\bar b H^0$ in the detection of the Higgs boson $H^0$
is more important than the case of two-body mode $(A^0, Z^0)+H^0$, because
$b\bar b H^0$ cover a major region in the parameters space $(m_{A^0}-\tan\beta)$.
The most important conclusion from this figure is that detection of all of the
neutral Higgs bosons will be possible at Next Linear $e^+e^-$ Collider.

\subsection{Detection of $A^0$}

For the pseudoscalar $A^0$, it is interesting to consider the production mode
in $b\bar b A^0$, since it can have large a cross-section due to the fact that
the coupling of $A^0$ with $b\bar b$ is directly proportional to $\tan\beta$,
thus will always grow with it. In Fig. 8, we present the contours of the
cross-sections for  the process of our interest $b\bar b A^0$, at $\sqrt{s}=500$
$GeV$ and ${\cal L}=50$ $fb^{-1}$.

We display the contour lines for $\sigma= 0.01, 0.003, 0.001$, showing also the
regions where the $A^0$ can be detected. These cross-sections give a contour of
production of 500, 150 and 50 events. It is clear from this figure that very high
experimental and analysis efficiencies are necessary for detecting the Higgs boson
$A^0$.

On the other hand, if we focus the detection of the $A^0$ at $\sqrt{s}=1$ $TeV$
and an integrated luminosity of ${\cal L} = 100$ $fb^{-1}$, the panorama for its
detection is more extensive. The Fig. 9 shows the contours lines in the plane
$(m_{A^0}-\tan\beta)$, to the cross-section of $b\bar bA^0$. The contours for this 
cross-section correspond to 300, 100 and 10 events. While for ${\cal L}=200$ $fb^{-1}$
we have 600, 200, and 20 events respectively. It is estimated that if more than 100
total events are obtained for $b\bar b A^0$, then it is possible to detect the $A^0$.

\subsection{Detection of $H^\pm$}

Our results for the $H^+H^-$ scalars are displayed in Fig. 10, 11 for $e^+e^-\to H^+H^-$
dominant mode and for the processes at three-body $e^+e^-\to \tau^-\bar \nu_\tau H^+,
\tau^+\nu_\tau H^-$.

The total cross-section for this reaction with $\sqrt{s}=500$ $GeV$ and
${\cal L}=50$ $fb^{-1}$ are shown in Fig. 10 for each contour with 0.01, 0.001,
and 0.0001 $pb$, which gives 500 events, 50 events, and 5 events, respectively.
We can see from this figure that the effect of the reactions $\tau^-\bar \nu_\tau H^+$
and $\tau^+\nu_\tau H^-$ is slightly more important than $H^+H^-$ for most of
the $(m_{A^0}-\tan\beta)$ parameters space regions. Nevertheless, there are
substantial portions of parameters space in which the discovery of the $H^\pm$
is not possible using either $H^+H^-$ or $\tau^-\bar\nu_\tau H^+$ and $\tau^+\nu_\tau H^-$.

In both cases the curves with values of 0.01 $pb$, 0.001 $pb$, and 0.0001 $pb$
give 1000, 100, and 10 events to an integrated luminosity of ${\cal L}=100$ $fb^{-1}$.
Meanwhile, for ${\cal L}=200$ $fb^{-1}$ we have 2000, 200, and 20 events. These
cross-sections are small, however, it is precisely in this curve where the contribution
of the processes at three-body is notable. The most important conclusion from
this figure is that detection of the charged Higgs bosons will be possible at
future $e^+e^-$ colliders.

\section{Conclusions}

In this paper, we have calculated the production of the neutral and charged Higgs
bosons in association with $b$-quarks and with $\tau \nu_\tau$ leptons via the
processes $e^{+}e^{-}\rightarrow b \bar b h$, $h = h^0, H^0, A^0$ and
$e^+e^-\to \tau^-\bar \nu_\tau H^+, \tau^+\nu_\tau H^-$ using the helicity
formalism. We find that these processes could help to detect the possible neutral
and charged Higgs bosons at energies of a possible Next Linear $e^+e^-$ Collider
when $\tan \beta$ is large.

In summary, we conclude that the possibilities of detecting or excluding the 
neutral and charged Higgs bosons of the Minimal Supersymmetric Standard Model
$(h^0, H^0, A^0, H^\pm)$ in the processes $e^{+}e^{-}\rightarrow b\bar b h$,
$h = h^0, H^0, A^0$ and $e^+e^-\to \tau^-\bar \nu_\tau H^+, \tau^+ \bar \nu_\tau H^-$
are important and in some cases are compared favorably with the dominant mode
$e^{+}e^{-}\rightarrow (A^0, Z^0) + h$, $h = h^0, H^0, A^0$ and $e^+e^-\to H^+H^-$
in the region of parameters space $(m_{A^0}-\tan \beta)$ with $\tan \beta$ large.
The detection of the Higgs boson will require the use of a future high energy
machine like the Next Linear $e^{+}e^{-}$ Collider.

\hspace{2cm}

\begin{center}
{\bf Acknowledgments}
\end{center}

This work was supported in part by {\it Consejo Nacional de Ciencia y
Tecnolog\'{\i}a} (CONACyT) (Proyecto I33022-E) and {\it Sistema Nacional de
Investigadores} (SNI) (M\'exico). A.G.R. would like to thank the organizers
of the Summer School in Particle Physics 99, Trieste Italy for their hospitality. 
O. A. S. would like to thank CONICET (Argentina).

\newpage

\begin{center}
{\bf FIGURE CAPTIONS}
\end{center}

\vspace{5mm}

\bigskip

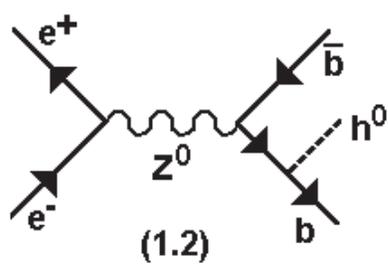
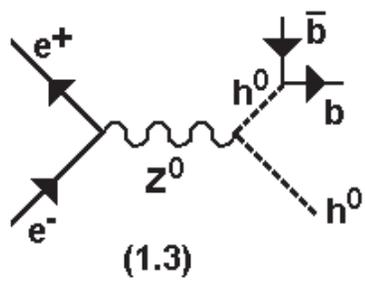
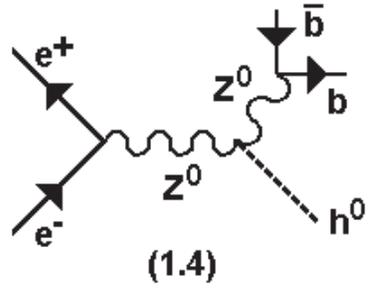
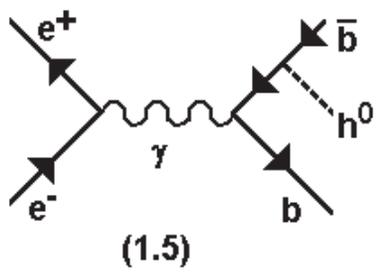
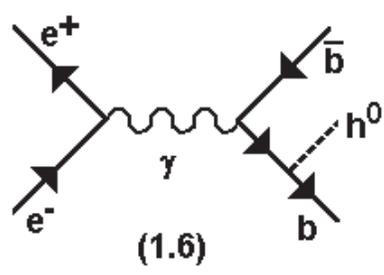
\noindent {\bf Fig. 1} Feynman Diagrams at tree-level for $e^{+}e^{-}
\rightarrow b\bar b h^0$. For $e^{+}e^{-}\rightarrow b\bar b H^0$
one has to make only the change $\sin\alpha / \cos\beta \rightarrow
\cos\alpha / \cos\beta$.

\bigskip

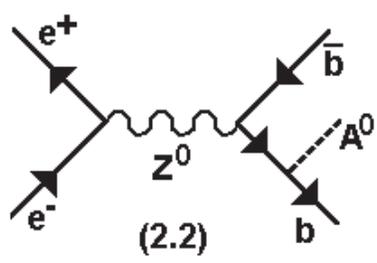
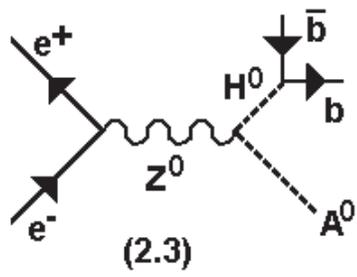
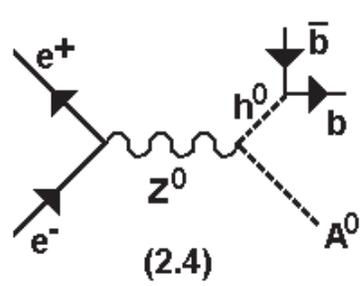
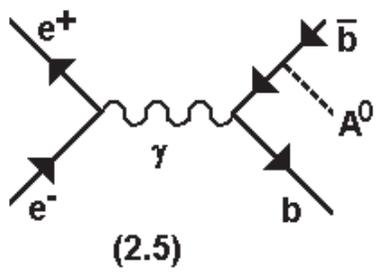
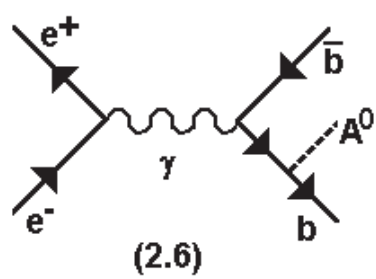
\noindent {\bf Fig. 2} Feynman Diagrams at tree-level for $e^{+}e^{-}
\rightarrow b\bar b A^0$.

\bigskip

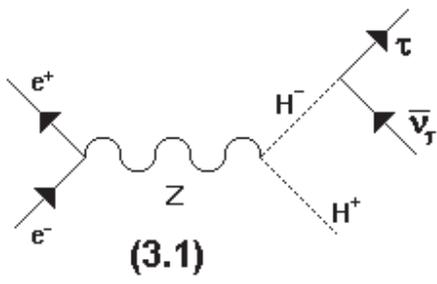
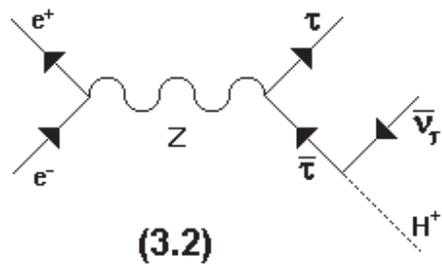
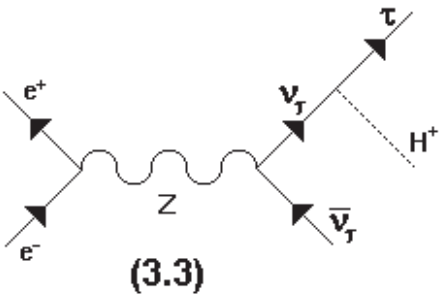
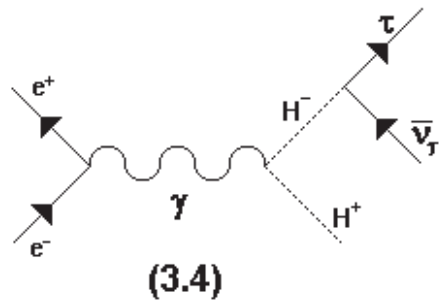
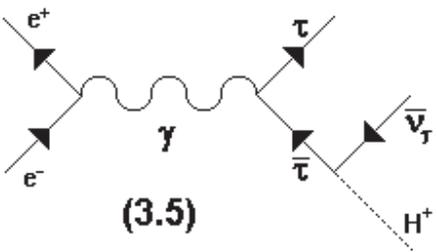
\noindent {\bf Fig. 3} Feynman Diagrams at tree-level for $e^{+}e^{-}
\rightarrow \tau^-\bar \nu_\tau H^+, \tau^+\nu_\tau H^-$.

\bigskip

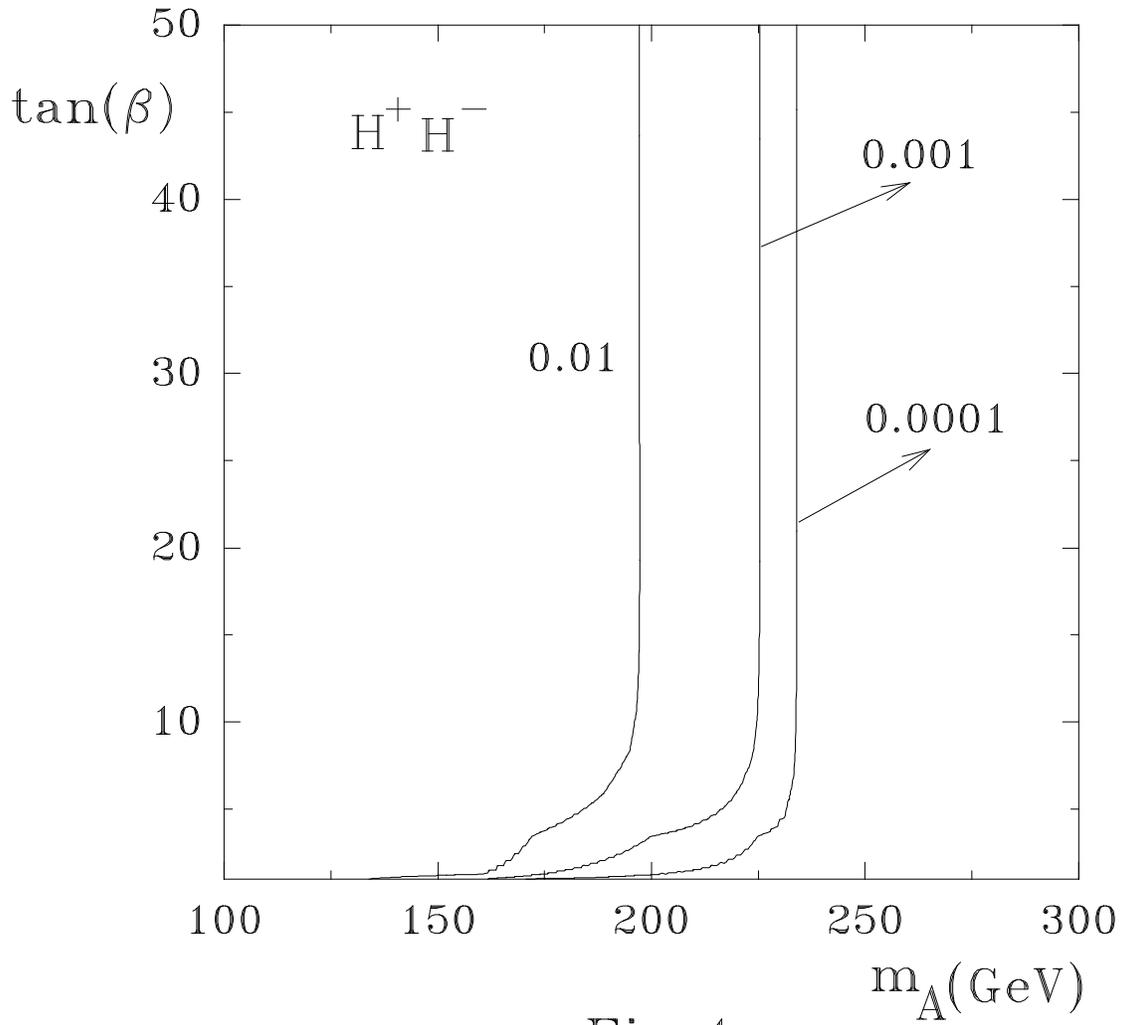
\noindent {\bf Fig. 4} Total cross-sections contours in $(m_{A^0}-
\tan\beta)$ parameter space for $e^{+}e^{-}\rightarrow (A^0, Z^0) + h^0$
and $e^{+}e^{-}\rightarrow b\bar b h^0$ with $\sqrt{s} = 500$
$GeV$ and an integrated luminosity of ${{\cal L} = 50}$ $fb^{-1}$. We have
taken $m_{t} = 175$ $GeV$ and $M_{\stackrel {\sim}t} = 500$ $GeV$ and neglected
squark mixing.

\bigskip

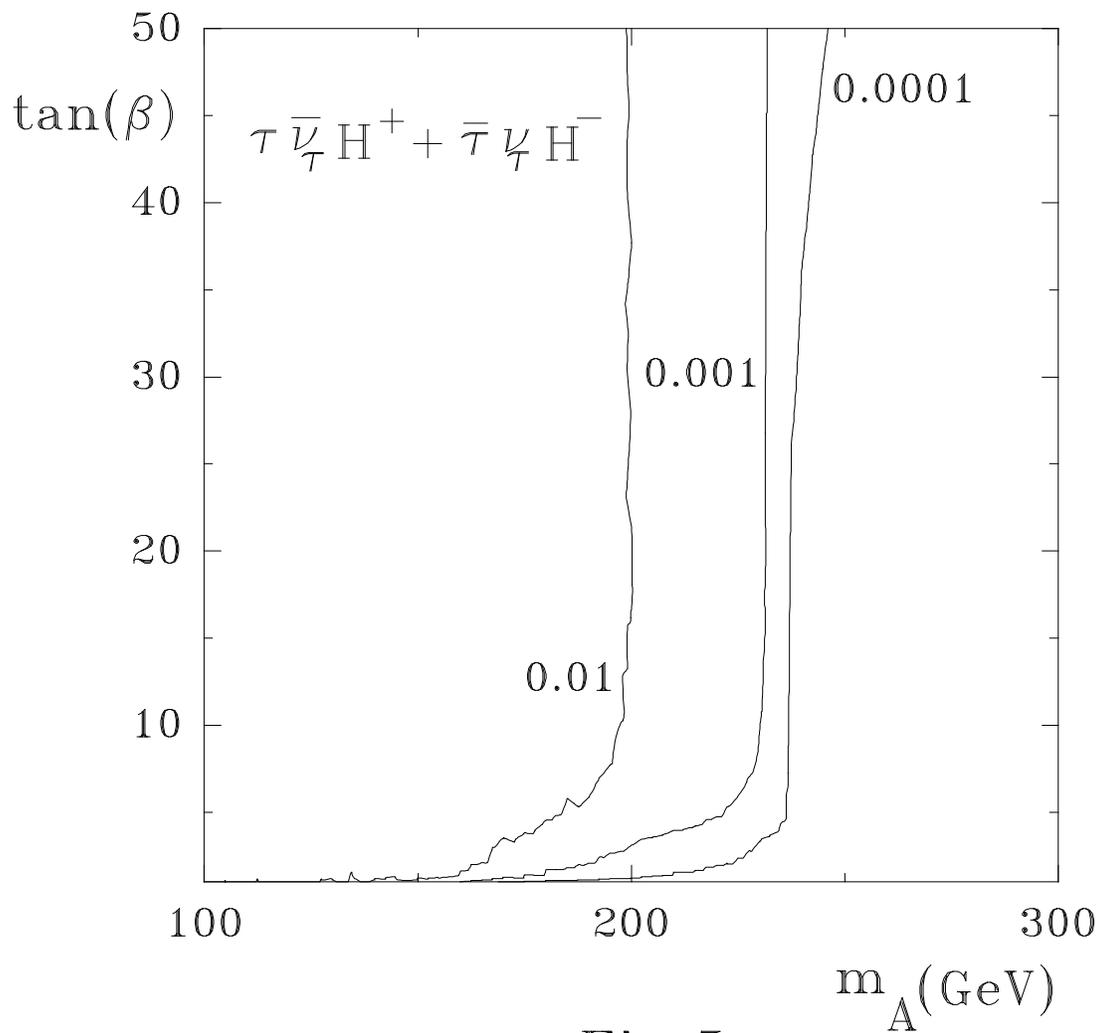
\noindent {\bf Fig. 5} Total cross-sections contours for 
$\sqrt{s}= 1$ $TeV$ and ${\cal L} = 100$, 200 $fb^{-1}$. We have taken
$m_{t} = 175$ $GeV$, $M_{\stackrel {\sim} t} = 500$ $GeV$ and neglected
squark mixing. We display contours for $e^{+}e^{-}\rightarrow (A^0, Z^0) +
h^0$ and $e^{+}e^{-}\rightarrow b\bar b h^0$, in the parameters space
$(m_{A^0}-\tan\beta)$.

\bigskip

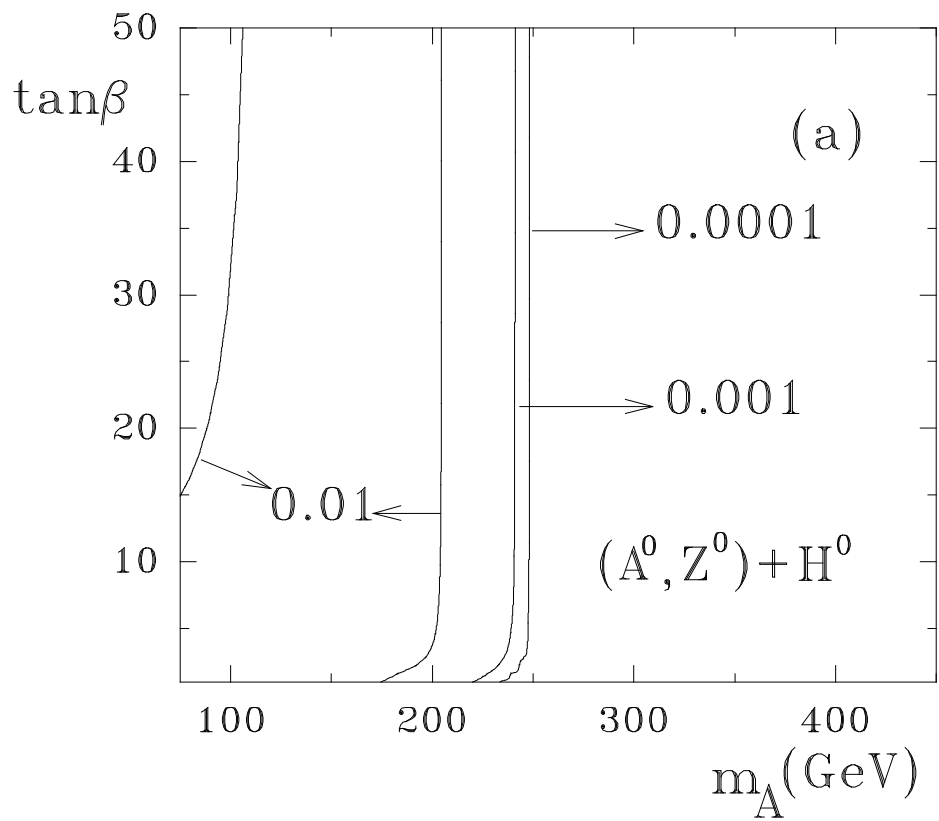
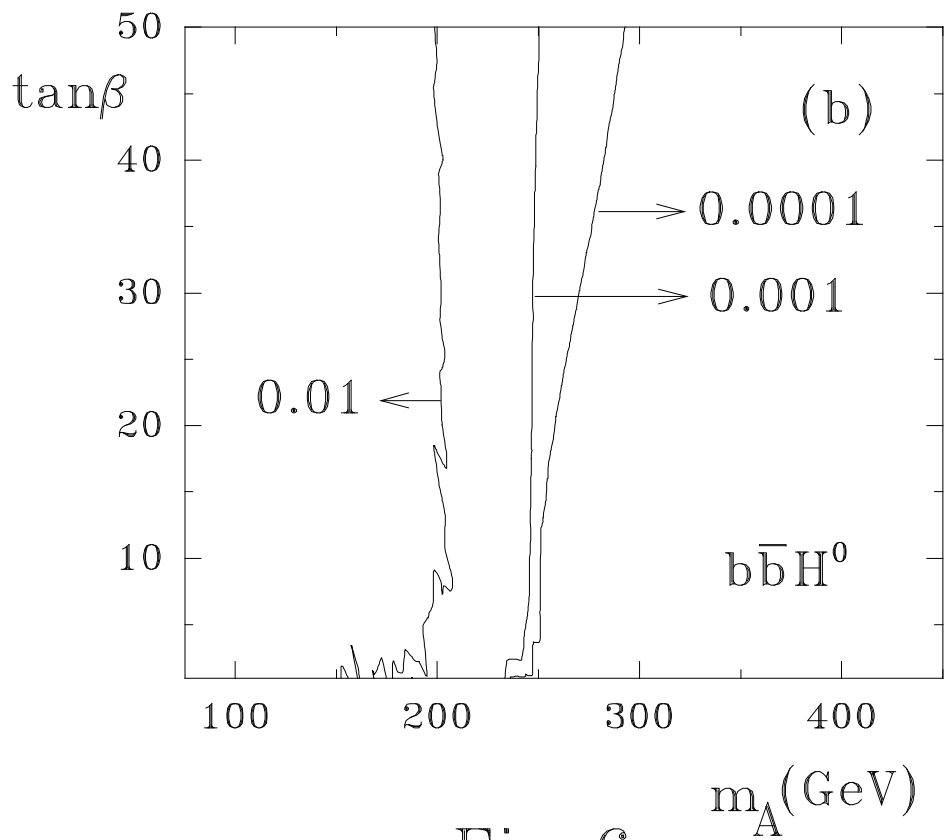
\noindent {\bf Fig. 6} Same as in Fig. 4, but for
$e^{+}e^{-}\rightarrow (A^0, Z^0) + H^0$ and $e^{+}e^{-}\rightarrow b
\bar b H^0$.

\bigskip

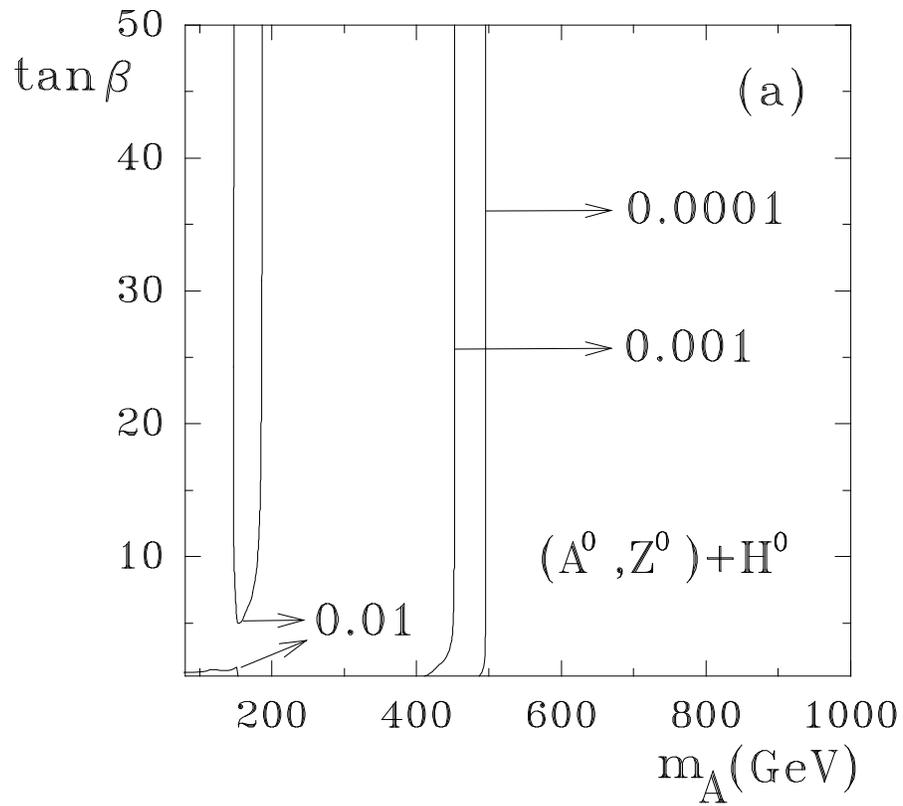
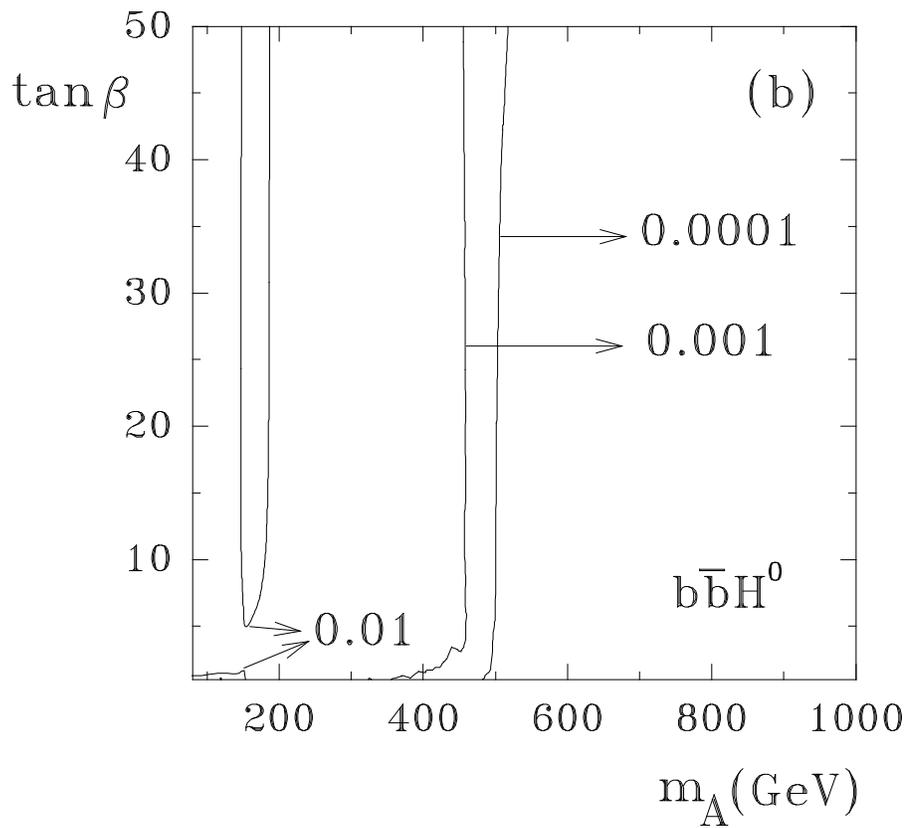
\noindent {\bf Fig. 7} Same as in Fig. 5, but for
$e^{+}e^{-}\rightarrow (A^0, Z^0) + H^0$ and $e^{+}e^{-}\rightarrow b
\bar b H^0$.

\bigskip

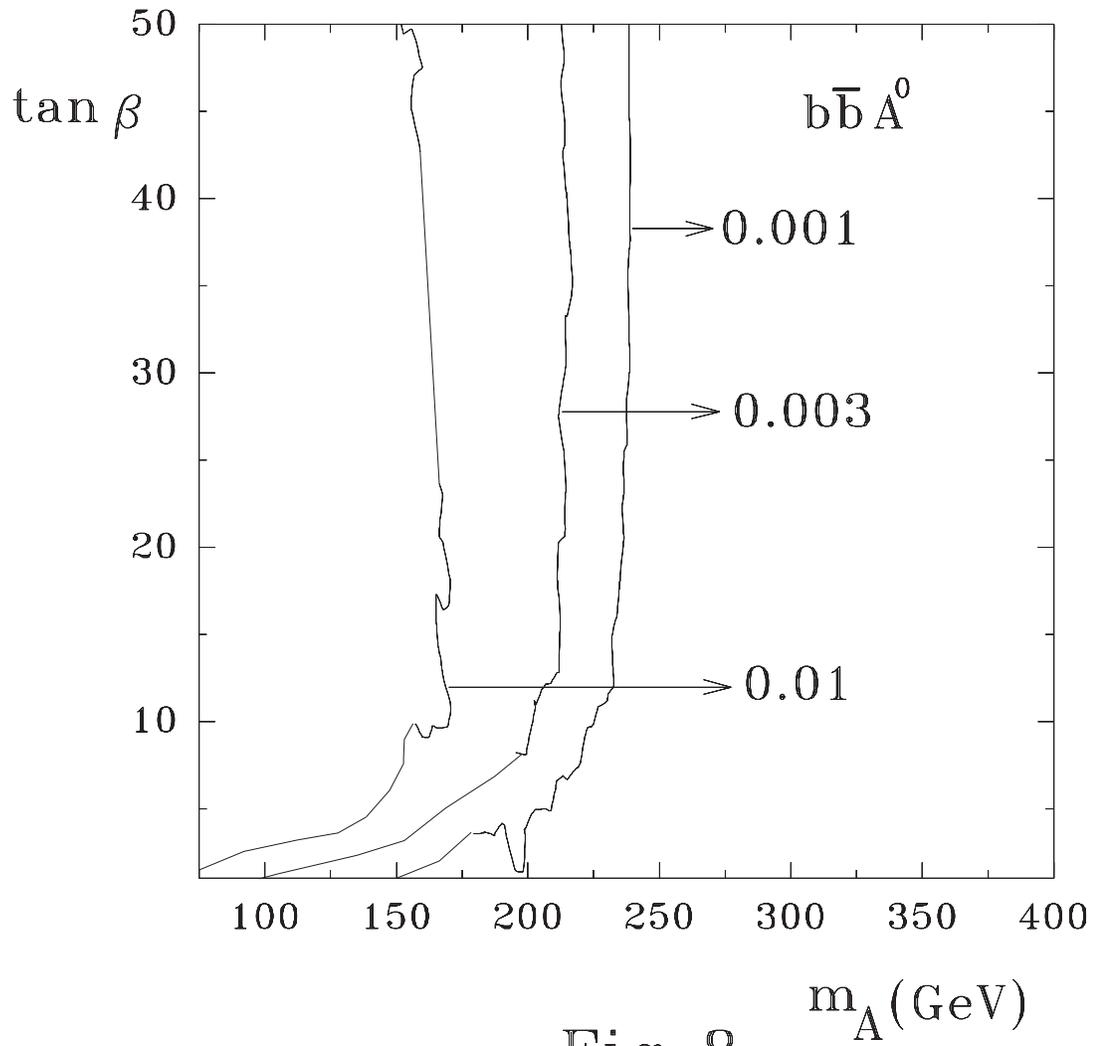
\noindent {\bf Fig. 8} Same as in Fig. 4, but for
$e^{+}e^{-}\rightarrow b\bar b A^0$.

\bigskip

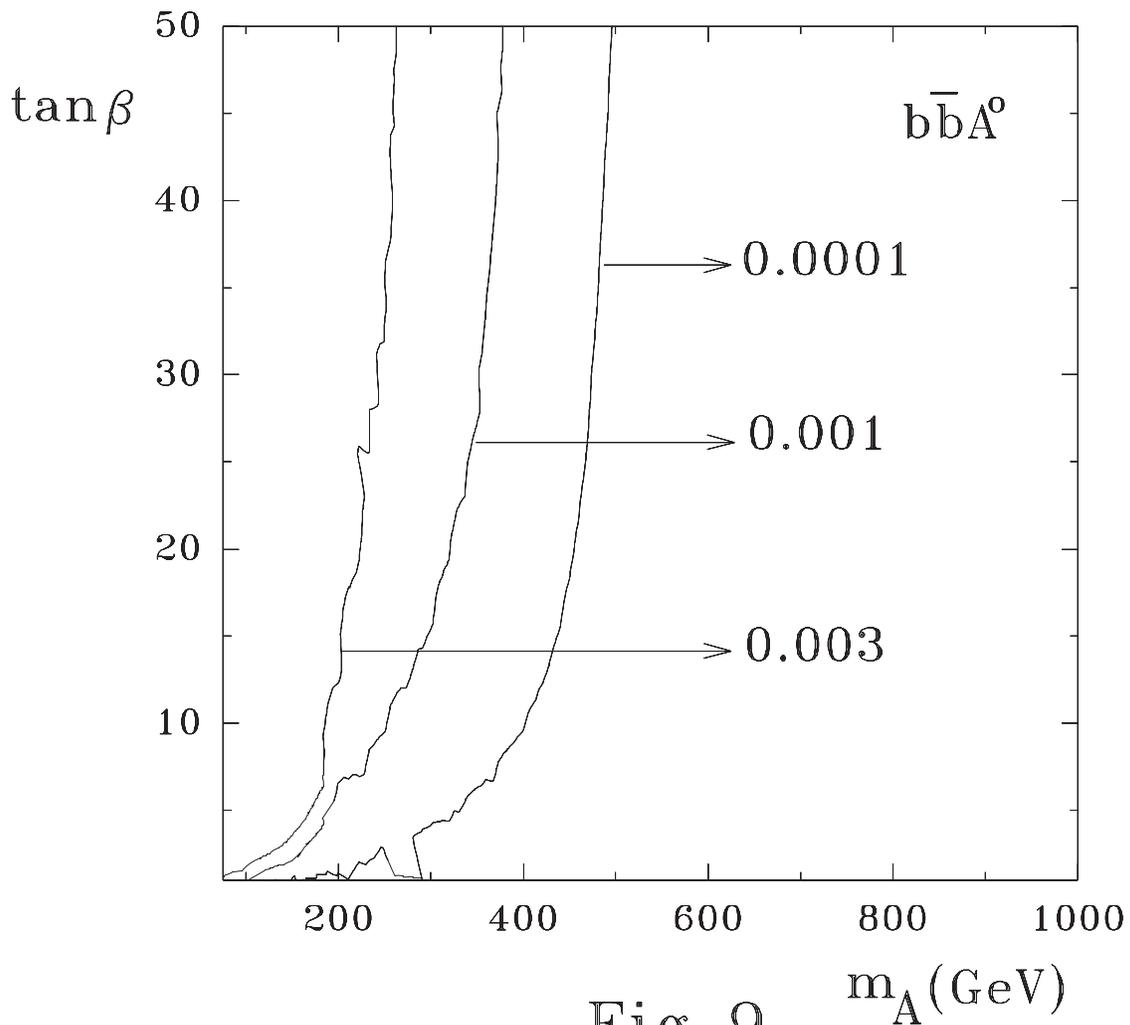
\noindent {\bf Fig. 9} Same as in Fig. 5, but for
$e^{+}e^{-}\rightarrow b \bar b A^0$.

\bigskip

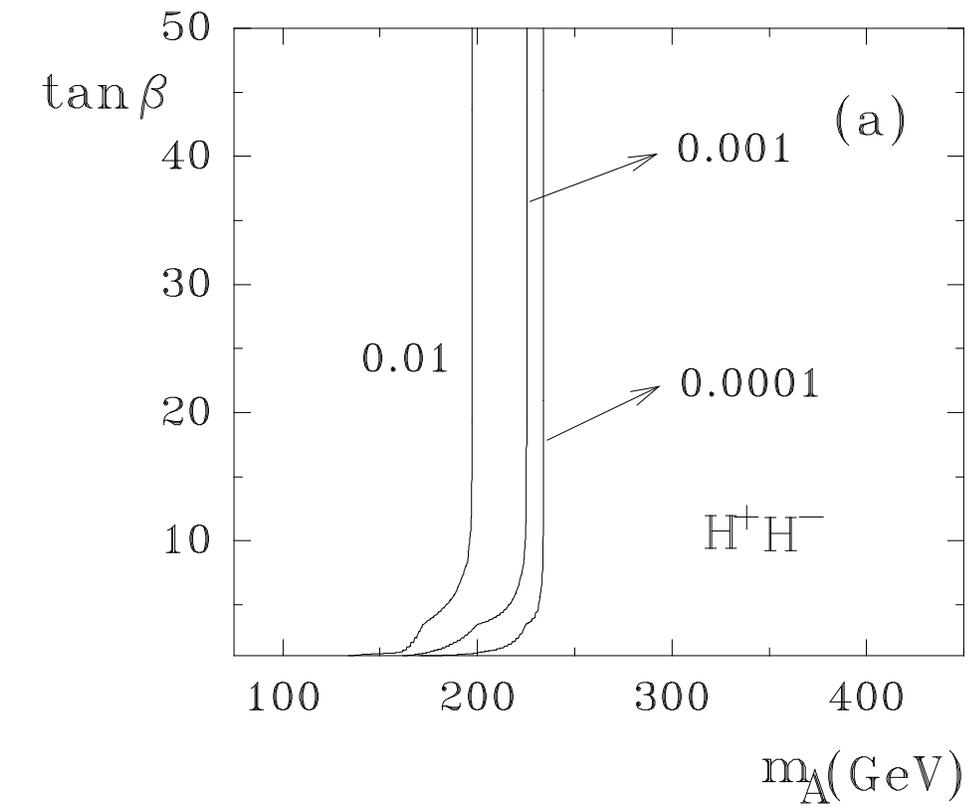
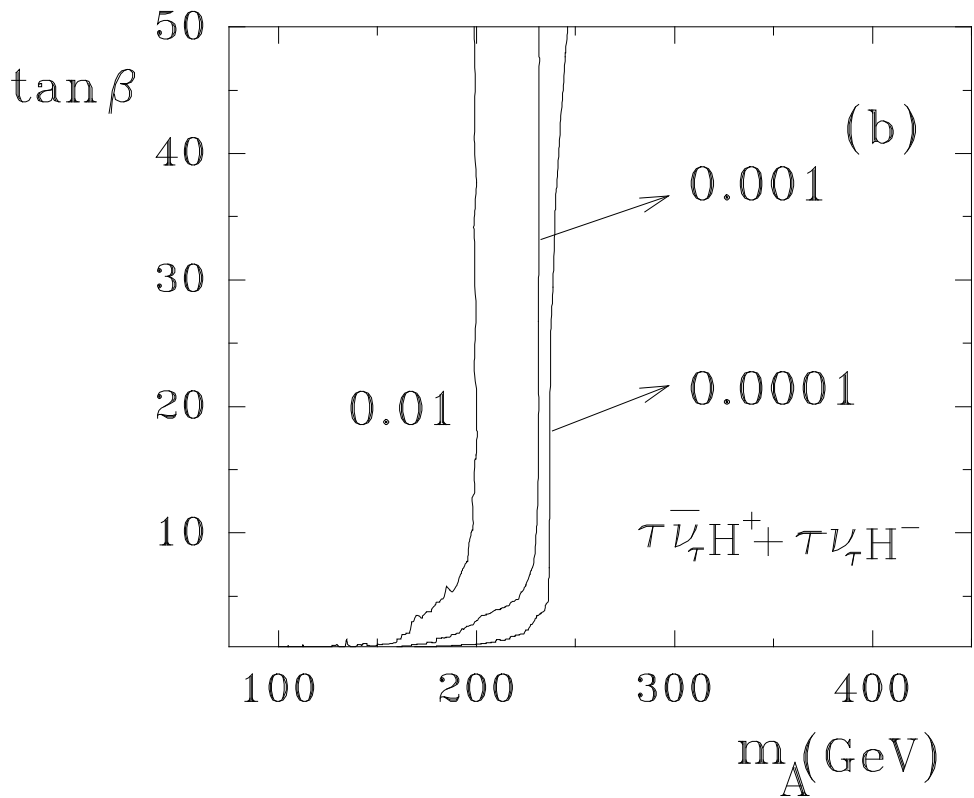
\noindent {\bf Fig. 10} Same as in Fig. 4, but for
$e^{+}e^{-}\rightarrow H^+ H^-$ and
$e^+e^-\to \tau^-\bar \nu_\tau H^+, \tau^+ \nu_\tau H^-$.

\bigskip

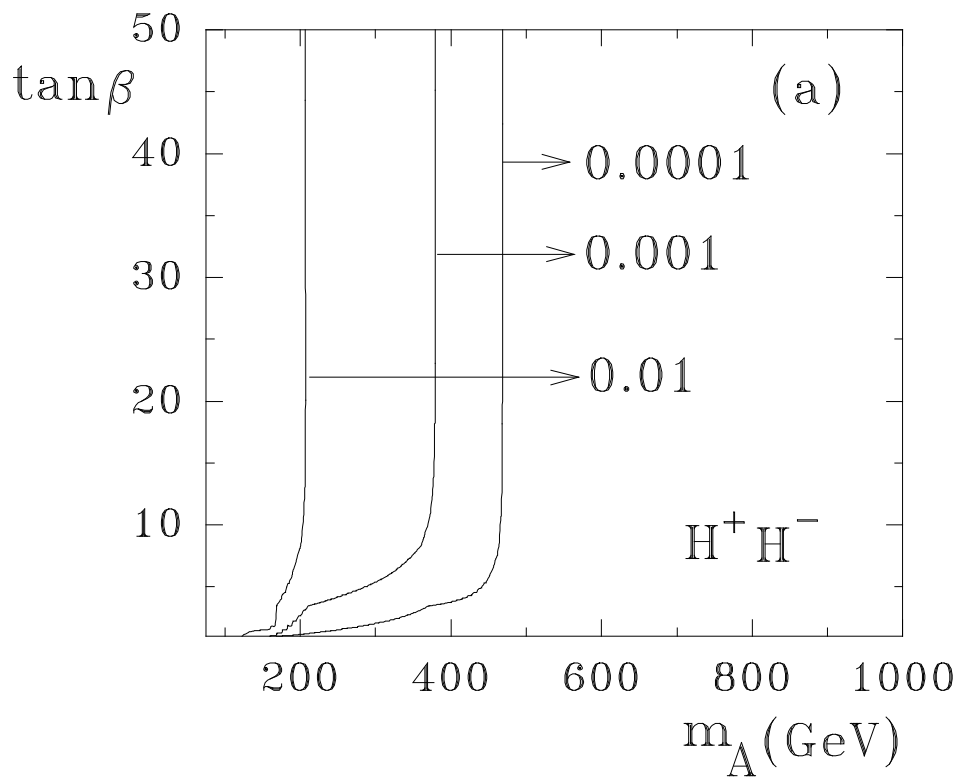
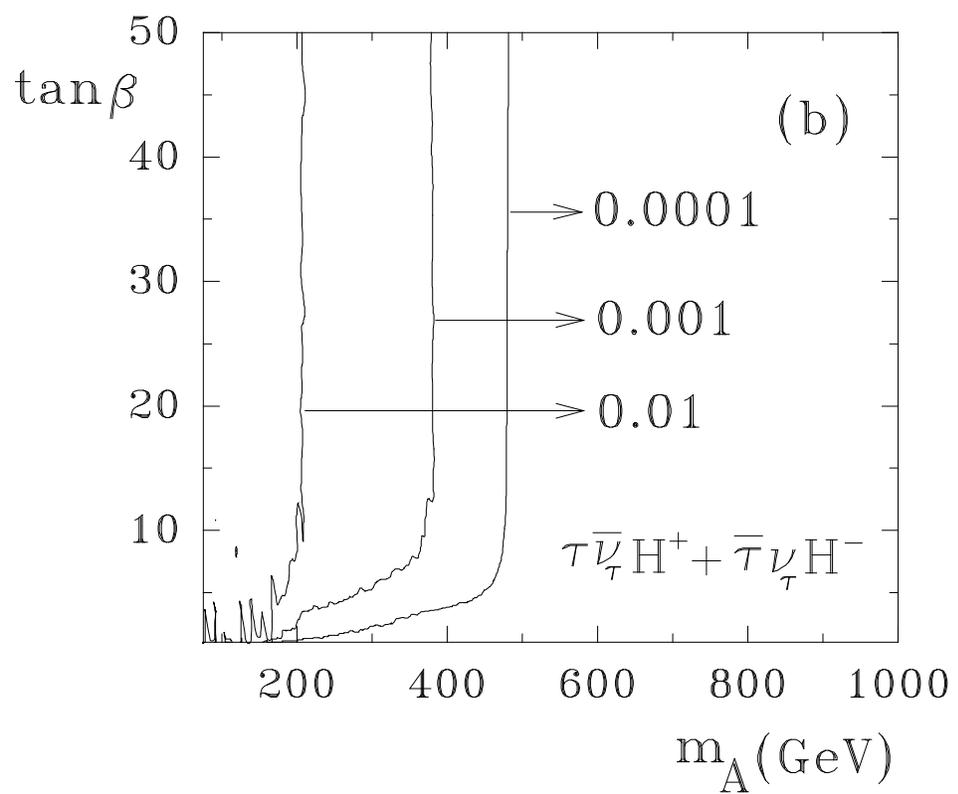
\noindent {\bf Fig. 11} Same as in Fig. 5, but for
$e^{+}e^{-}\rightarrow H^+ H^-$ and
$e^+e^-\to \tau^-\bar \nu_\tau H^+, \tau^+ \nu_\tau H^-$.

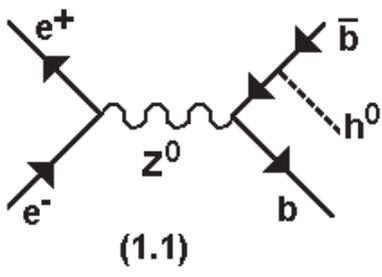
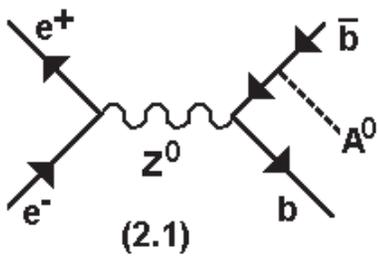
\end{document}